\documentclass[twocolumn,prl,aps,superscriptaddress,showpacs]{revtex4}
\usepackage{graphicx}

\newcommand{\beq}{\begin{equation}}
\newcommand{\eeq}{\end{equation}}

\begin{document}

\title{Generalized fluctuation relation and effective temperatures
in a driven fluid
}
\author{F.~Zamponi}
\affiliation{Dipartimento di Fisica and INFM, Universit\`a di Roma
    {\em La Sapienza}, P. A. Moro 2, 00185 Roma, Italy}
\author{G.~Ruocco}
\affiliation{Dipartimento di Fisica and INFM, Universit\`a di Roma
    {\em La Sapienza}, P. A. Moro 2, 00185 Roma, Italy}
\affiliation{INFM - CRS {\it Soft},
    Universit\`a di Roma {\em La Sapienza}, P. A. Moro 2, 00185 Roma,
    Italy}
\author{L.~Angelani}
\affiliation{Dipartimento di Fisica and INFM, Universit\`a di Roma
    {\em La Sapienza}, P. A. Moro 2, 00185 Roma, Italy}
\affiliation{INFM - CRS {\it SMC},
      Universit\`a di Roma {\em La Sapienza}, P. A. Moro 2, 00185 Roma,
      Italy}
\date{\today}
\begin{abstract}
By numerical simulation of a Lennard-Jones like liquid
driven by a velocity gradient $\gamma$
we test the fluctuation relation (FR)
below the (numerical) glass transition temperature $T_g$.
We show that, in this region,
the FR deserves to be
generalized introducing a numerical factor $X(T,\gamma)<1$
that defines an ``effective temperature'' $T_{FR}=T/X$.
On the same system we also measure the effective temperature 
$T_{eff}$, as defined from the generalized
fluctuation-dissipation relation, and find a qualitative agreement
between the two different nonequilibrium temperatures.
\end{abstract}
\maketitle


The {\it Fluctuation Theorem} (FT) concerns
the fluctuations of the entropy production rate $\sigma(t)$ 
in the stationary nonequilibrium states of a {\it chaotic} driven system.
We will set $k_B =1$ and define 
$\sigma_+ \equiv \langle \sigma(t) \rangle$,
where $\langle \cdot \rangle$ is the time average in stationary
state; if $\sigma_+ > 0$ the system is out of equilibrium.
Defining the variable $p(t) = (\tau\sigma_+)^{-1} \int_t^{t+\tau} ds \, \sigma(s)$
(such that $\langle p \rangle$$=$$1$),
its Probability Distribution Function (PDF) $\pi_\tau(p)$,
and the large deviations function
\beq
\label{zetap}
\zeta_\tau(p)= \tau^{-1} \log \pi_\tau(p) \, , \hspace{20pt}
\zeta(p) = \lim_{\tau\rightarrow\infty} \zeta_\tau(p) \ ,
\eeq
the FT states that, for $|p|<p^*$ 
(where $p^*$ is defined by $\lim_{p\rightarrow \pm p^*} \zeta(p)=-\infty$),
the following relation --also called {\it Fluctuation Relation} (FR)-- must hold:
\beq
\label{FT}
\zeta(p)-\zeta(-p) = p\sigma_+ \ .
\eeq
The validity of this relation was first shown by 
Evans {\it et al} in a numerical simulation of a sheared fluid \cite{EGD}
and subsequently proven for {\it reversible} Anosov systems by Gallavotti and Cohen
\cite{GC}. Gallavotti then showed that, close to equilibrium ($\sigma_+ \rightarrow 0$), 
the FR implies the usual Fluctuation-Dissipation Relation (FDR) \cite{GAL}.
In the recent past, the FR has been tested under a wide class
of different conditions, and is now believed to be a very general relation 
for chaotic systems \cite{BG,FTgen,Kurchan,noiJSP}; 
recently, it has been also tested in some experiments \cite{FTexp}.

Extending the FT to the case of driven Langevin systems,
Kurchan pointed out that ``the FR might be
violated for those (infinite) driven systems which in the absence of drive
have a slow relaxational dynamics that does not lead them to equilibrium
in finite times'' \cite{Kurchan}. This is (by definition) the case of driven
glassy systems.
Driven glassy systems have been widely studied by numerical 
simulations: in \cite{drivenglassy}
a uniform velocity gradient $\gamma$ was applied 
on a Lennard-Jones liquid (which manifests glassy behavior below the 
{\it glass transition temperature} $T_g$ \cite{nota1}) 
at fixed kinetic temperature $T$.
In presence of the driving force, the system becomes stationary also below $T_g$,
while in the absence of drive the system is not able to equilibrate with the bath and
{\it ages} indefinitely.
It was shown that below $T_g$
the FDR does not hold anymore
(because in absence of drive
the system is not able to reach equilibrium below $T_g$) 
but can be generalized, for small driving force,
introducing an ``effective temperature'' $T_{eff}$ --higher than the temperature
of the bath--
associated with the ``slow'' modes that in absence of drive are responsible 
for the glassy behavior \cite{FDTshear,drivenglassy}.
The breakdown of the FDR and the close
relation between the latter and the FR
support the conjecture of Kurchan that the FR also has to be modified
below $T_g$.

A possible generalization of the FR, of the form
\beq
\label{FTX}
\zeta(p)-\zeta(-p) = X p \sigma_+ \ ,
\eeq
was proposed in \cite{BG}
in the context of chaotic dynamical systems.
It has also been proposed to define $T_{FR}\equiv T/X$ as the ``temperature''
in nonequilibrium steady states \cite{gal1}.
A similar generalization has been proposed by many authors in the context of glassy 
systems, following the reported Kurchan's observation, and some attempts
have been made in order to relate $T_{FR}$ with the effective temperature
$T_{eff}$ introduced in the generalized FDR \cite{FRgen-vetri,semerjian}.
Recently, a connection between the generalized FDR and the FR has been derived
in a model for the Brownian diffusion of a particle in a nonequilibrium environment
\cite{CKZ04}.

However, up to now numerical studies of the FR have been
performed only in the high temperature region ($T \gg T_g$).
The aim of this paper is to test the FR {\it below} $T_g$ 
in a numerical simulation of a Lennard-Jones like liquid.
We measured $\zeta(p)$ and found that the data are 
consistent with Eq.~\ref{FTX} with $X<1$, while above $T_g$ one 
has $X=1$, consistently with what has been found in previous works.
We measured also the effective temperature $T_{eff}$ from the generalized
FDR and found a good agreement between $T_{FR}=T/X$ and $T_{eff}$.

The investigated system is a 80:20 binary mixture of 
$N$=66 particles of equal mass $m$ interacting via a Soft Sphere Potential (SSP)
$V_{\alpha \beta}(r)=\epsilon_{\alpha \beta} 
\left( \frac{\sigma_{\alpha\beta}}{r} \right)^{12}$ ($\alpha,\beta \in [A,B]$).
This system has been introduced and characterized in equilibrium by De Michele {\it et al}
\cite{PhDcristiano}
as a modification of the standard LJ Kob-Andersen mixture that is known to avoid
crystallization on very long time scales, and hence to be a very good model of glass former;
it has been chosen because the SSP can be cut at very short distance ($\sim 1.5 \sigma_{AA}$) allowing
the system to be very small ($N$=66) in order to observe the negative values of $p$ that
are required to test Eq.~\ref{FT}.
All the quantities are reported in units of
$m$, $\epsilon_{AA}$ and $\sigma_{AA}$.
In these units the integration step is $dt=0.005$.
The particles are confined in a cubic box with Lees-Edwards boundary conditions \cite{evans} at
density $\rho = 1.2$.
 The shear flow is applied to the system
along the $x$ direction with a gradient velocity field along the
$y$ axis. The molecular dynamics simulation is performed using
SLLOD equations of motion \cite{evans}:
\beq
\dot{q}_i = \frac{p_i}{m} + \gamma q_{yi} \hat{x} \, , \hspace{1cm}
\dot{p}_i = F_i(q) - \gamma p_{yi} \hat{x} - \alpha p_i \, ,
\eeq
where $F_i(q)=-\partial_{q_i} V(q)$ and 
$\alpha$ is a Gaussian thermostat that fixes the 
kinetic energy $\sum_i \frac{p_i^2}{2m} = \frac{3}{2}NT$.
The entropy production rate is defined as the
dissipated power $W$ divided by the kinetic temperature $T$ \cite{nota2}:
$\sigma(p,q)=W(p,q)/T=-\gamma P_{xy}(p,q) / T$,
where $P_{xy}(p,q)=\sum_i [ p_{xi} p_{yi} + q_{yi} F_{xi}(q) ]$ is the $xy$
component of the stress tensor \cite{evans}.

\begin{figure}[ht]
\centering
\includegraphics[width=.45\textwidth,angle=0]{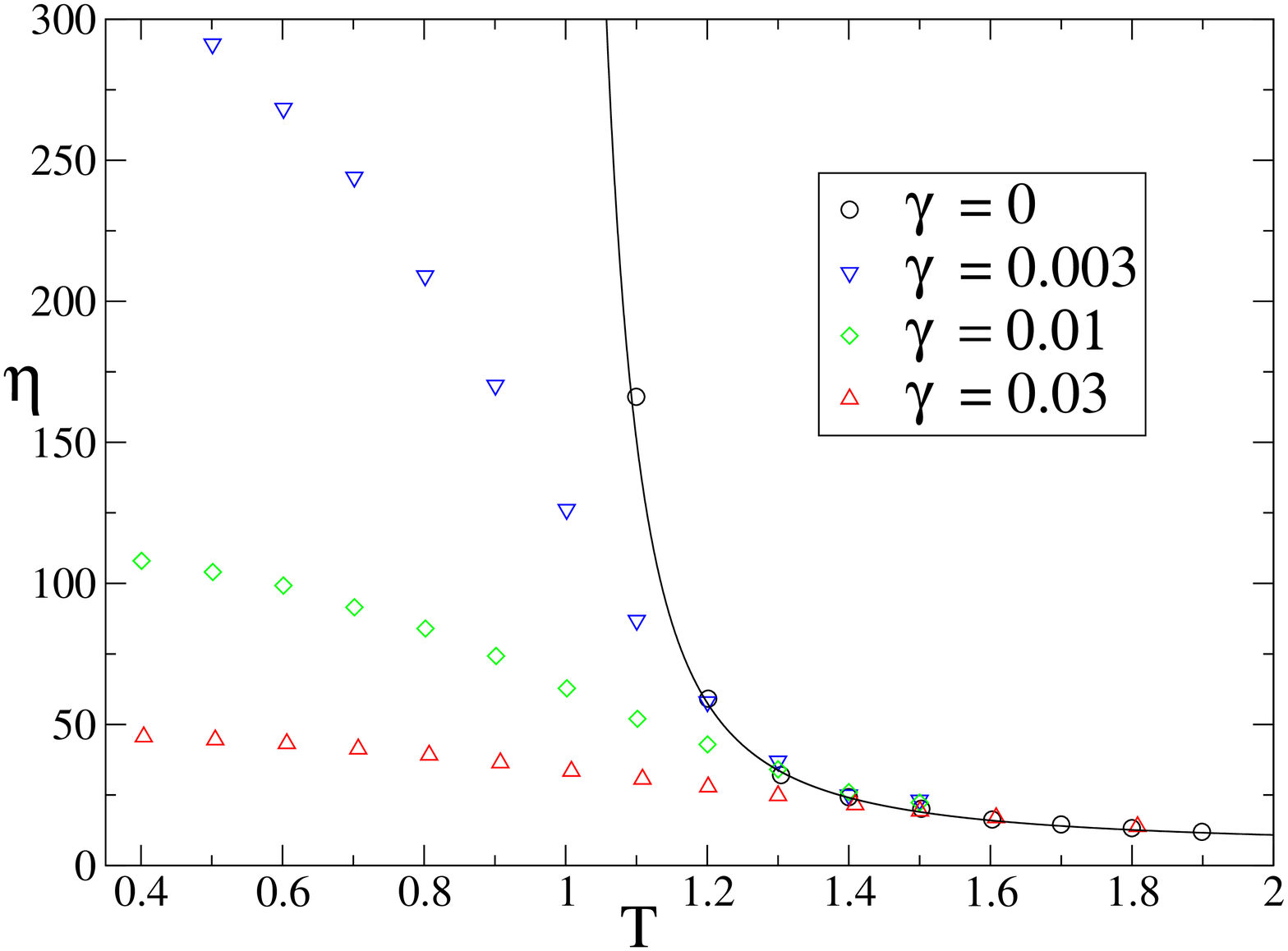}
\caption{Viscosity as a function of temperature for different values of $\gamma$. 
The continuous line is a fit to a Vogel-Tamman-Fulcher law,
$\eta(T)=\eta_0 \exp (\frac{AT_0}{T-T_0})$ with $\eta_0=5.2$, $A=0.99$, $T_0=0.85$. 
}
\label{fig_1}
\end{figure}

In Fig.~\ref{fig_1} we report the viscosity $\eta \equiv \langle P_{xy} \rangle /\gamma$
as a function of the temperature $T$ for different values of the shear rate $\gamma$.
At $\gamma=0$ the viscosity seems to diverge at a temperature $T_0 \sim 0.85$;
however, we are able to equilibrate our system only down to $T \sim 1.1$, that provides
an estimate for the glass transition temperature $T_g$.
For $\gamma > 0$ the system becomes stationary and the viscosity is finite 
at all temperatures, even below $T_0$.

Very long simulation runs (up to $2 \cdot 10^9$ time steps) have been performed
to measure the PDF of the entropy production rate along the line $\gamma=0.03$.
During the run, $p(t)$ has been measured on subsequent time intervals of 
duration $\tau$. 
From this dataset, we constructed the histograms of $\pi_\tau(p)$ and
the large deviations function $\zeta_\tau(p)$ defined in Eq.~\ref{zetap}.
The function $\zeta_\tau(p)$ is observed to converge to its asymptotic value
$\zeta(p)$ for $\tau \gtrsim \tau_\alpha$, $\tau_\alpha$
being the relaxation time of
the autocorrelation function of $\sigma(t)$.

\begin{figure}[t]
\centering
\includegraphics[width=.50\textwidth,angle=0]{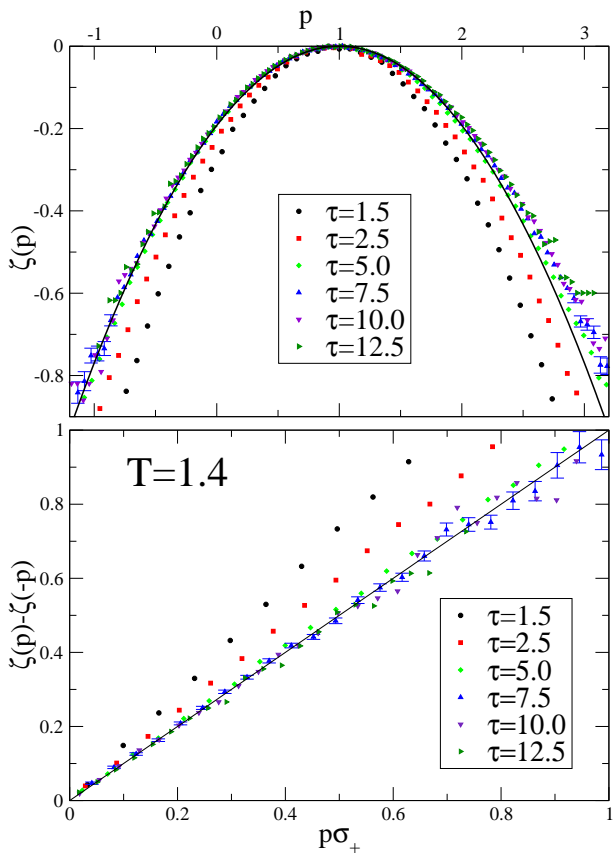}
\caption{Top: the large deviation function 
$\zeta_\tau(p)=\tau^{-1} \log \pi_\tau(p)$ as a function of 
$p$ for different values of $\tau$ at 
$T=1.4 > T_g$ and $\gamma=0.03$.
Error bars are smaller than the symbols except on the tails: they are reported
only for $\tau=7.5$ to avoid confusion.
The line is a Gaussian fit to the data with $\tau > 5$ for $p \in [0,2]$.
Bottom: $\zeta_\tau(p)-\zeta_\tau(-p)$ as a function of $p\sigma_+$.
The FR predicts the plot to be a straight line with slope 1 (full line)
for large $\tau$.
}
\label{fig_2}
\end{figure}

In the upper panel of Fig.~\ref{fig_2}, we report the functions
$\zeta_\tau(p)$ for $\gamma=0.03$ and 
$T=1.4 > T_g$.
The asymptotic function $\zeta(p)$ is obtained for $\tau \gtrsim 5$
and can be described by a simple Gaussian form, 
$\zeta(p) = -(p-1)^2/2\delta^2$, even if small non-Gaussian tails are observed.
In the lower panel of Fig.~\ref{fig_2} we report
$\zeta_\tau(p)-\zeta_\tau(-p)$ as a function of $p\sigma_+$.
The FR, Eq.~\ref{FT}, predicts the plot to be a straight line with slope 
1 for large $\tau$; this is indeed the case for $\tau \gtrsim 5$,
consistently with what has been found in the literature~\cite{EGD,FTgen}.

\begin{figure}[t]
\centering
\includegraphics[width=.50\textwidth,angle=0]{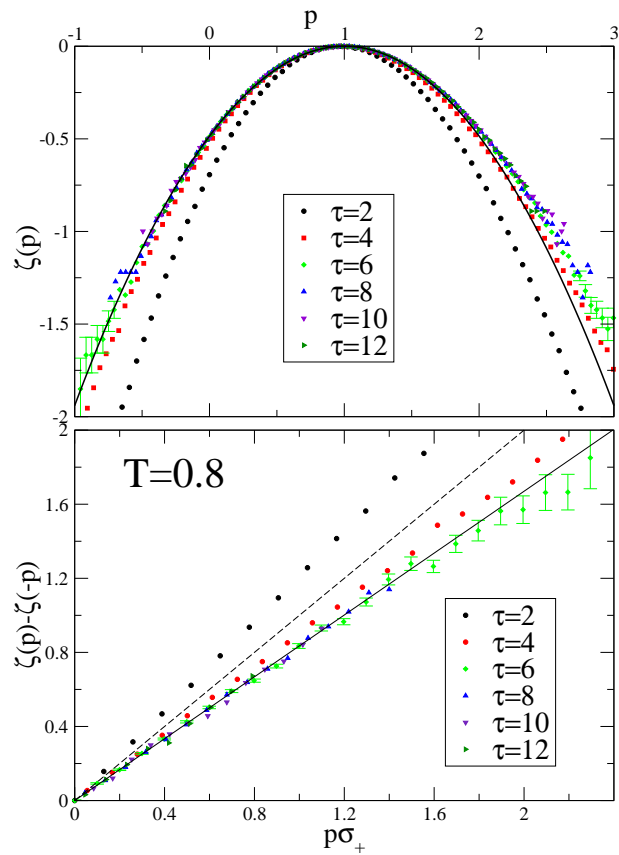}
\caption{Same plots as in Fig.~\ref{fig_2} for $\gamma=0.03$ and 
$T=0.8 < T_g$. In the lower panel the dashed line has slope
1 while the full line has slope $X = 0.83$.}
\label{fig_3}
\end{figure}

In the upper panel of Fig.~\ref{fig_3}, we report the functions $\zeta_\tau(p)$ for
$\gamma=0.03$ and $T=0.8 < T_g$. In this case, the asymptotic
regime is reached for $\tau \gtrsim 6$; this value is not so different from
the one obtained in the previous case because the change in viscosity (and hence
in relaxation time) going from $T=1.4$ to $T=0.8$ is very small at this
value of $\gamma$
(see Fig.~\ref{fig_1}).
Also in this case the simple Gaussian form gives a good description 
of the data apart from the small non-Gaussian tails.
In the lower panel of Fig.~\ref{fig_3} we report
$\zeta_\tau(p)-\zeta_\tau(-p)$ as a function of $p \sigma_+$.
At variance to what happens for $T > T_g$, in this case the asymptotic
slope reached for $\tau \gtrsim 6$ is smaller than 1;
thus, the FR given by Eq.~\ref{FT} has to be 
generalized according to Eq.~\ref{FTX}.
At this temperature, we get $X = 0.83 \pm 0.05$.

\begin{figure}[t]
\centering
\includegraphics[width=.45\textwidth,angle=0]{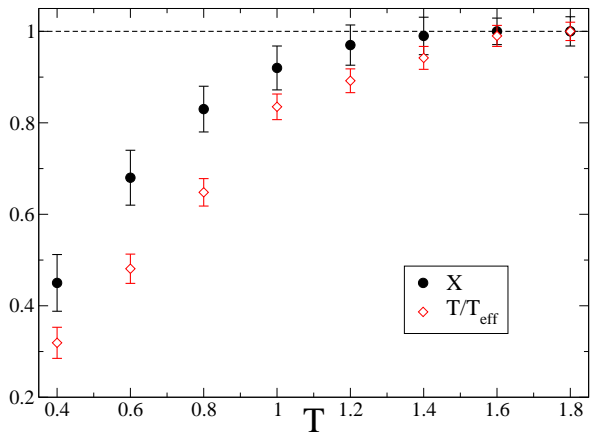}
\caption{The violation factor $X=T/T_{FR}$ that enters Eq.~\ref{FTX} (full circles)
and the ratio $T/T_{eff}$ from the generalized FDR (open diamonds)
as a function of the bath temperature $T$ for $\gamma=0.03$.}
\label{fig_4}
\end{figure}

In Fig.~\ref{fig_4}, we report the behavior of the violation factor 
$X(T,\gamma=0.03)$ (full circles) as a function of the temperature $T$; note that 
$X$ becomes smaller than unity exactly around $T_g \sim 1.1$, {\it i.e.} when the
viscosity starts to diverge strongly (see Fig.~\ref{fig_1}).
Below $T \sim 0.4$, $\sigma_+$ becomes so large that negative
fluctuations of $p$ are extremely rare and the violation factor is no longer
measurable. We can conclude that below $T_g$ the FR does not hold, and our data
are consistent with Eq.~\ref{FTX} where the coefficient
$X$ is temperature dependent below $T_g$ and equals $1$ above $T_g$.

Having checked the validity of Eq.~\ref{FTX}, following Ref.~\cite{gal1}, 
we can define a nonequilibrium temperature as $T_{FR}=T/X$. 
Note that if we define $\sigma'(t) = W/T_{FR} = X \sigma(t)$, 
the variable $p$ and the function $\zeta(p)$ are not affected by the 
rescaling while $\sigma'_+ = X \sigma_+$.
Thus, the FR for $\sigma'$ is the usual one given by Eq.~\ref{FT}. 
If one assumes that the fluctuations of entropy production rate should
satisfy Eq.~\ref{FT}, the entropy production rate should be given by
$\sigma'(t)$ instead of $\sigma(t)$.

It is interesting to compare
the temperature $T_{FR}$ with the effective temperature
$T_{eff}$ that enters the generalized Fluctuation Dissipation Relation.
The latter can be measured from the relation $T_{eff}=D/\mu$,
where $D$ is the diffusion constant and 
$\mu$ is the mobility of the particles in the considered steady state 
\cite{drivenglassy,FDT-MSD}.
This relation generalizes the usual equilibrium FDR $D=\mu T$;
to compute the diffusion constant and the mobility of type-A particles
we followed the procedure discussed in Ref.~\cite{FDT-MSD}.
In Fig.~\ref{fig_4}, together with $X=T/T_{FR}$, we report the ratio $T/T_{eff}$ 
(open diamonds) as a function of the bath temperature $T$.
The two ``effective'' temperatures have a similar qualitative behavior but do not coincide.

The origin of this discrepancy will be discussed in detail in \cite{CKZ04};
roughly speaking, the point is that the modes at all frequencies contribute
to the entropy production rate, while $T_{eff}$ is the temperature of the
slowest modes in the systems. At the values of $\gamma$ we considered, the
separation between a ``fast'' and a ``slow'' relaxation is not so sharp:
hence, $T_{FR}$ should be related to an average
over all the frequencies of the frequency-dependent effective temperature.
Note also that we are forced to use a very small system in order to observe
large fluctuations of the entropy production rate, thus, size effects
could affect the behavior of the investigated
quantities. We believe however that the qualitative picture is correct
even if size effects are not completely negligible.
Future works will hopefully clarify this issue
by exploring lower values of $\gamma$ for which separation of time scales is
more marked; however, for low values of $\gamma$, size effects are more
relevant and the dynamics of the system is very slow; thus, very long
simulations of bigger systems, requiring a large amount of CPU
time, are mandatory.

\begin{figure}[t]
\centering
\includegraphics[width=.45\textwidth,angle=0]{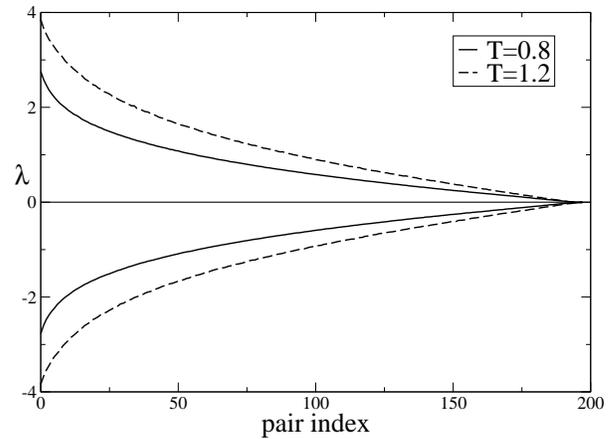}
\caption{
Lyapunov exponents for $\gamma=0.03$ and $T=0.8,1.2$. For both temperatures
each pair consists of one positive and one negative exponent, thus ${\cal D/N}=1$.
}
\label{fig_5}
\end{figure}

An interesting microscopical derivation of Eq.~\ref{FTX} was proposed by
Bonetto and Gallavotti \cite{BG}, who
related the factor $X$ to the dimensionality of the attractive set of the 
system in its phase space.
The latter can be measured by computing the Lyapunov spectrum,
which in this kind of system is composed by {\it pairs} of conjugated
exponents; the latter are constructed by pairing the largest exponent with
the smallest one and so on~\cite{pairing}.
The prediction of \cite{BG} is that $X={\cal D/N}$, where ${\cal D}$ is the number of 
pairs where one exponent is positive and the other is negative, 
and ${\cal N}$ is the total number of pairs.
If the attractor is dense in phase space, ${\cal D}={\cal N}$ and $X=1$.
This relation is very interesting as --if true-- it provides a link
between the effective temperature and properties of 
the phase space of the system.
The Lyapunov spectra have been measured by the mean of the standard 
algorithm of Benettin {\it et al} \cite{BGS76} and are reported in Fig.~\ref{fig_5}
for $\gamma=0.03$, $T=1.2> T_g$ and $T=0.8<T_g$. 
Unfortunately, no qualitative change in the spectrum 
is observed on crossing $T_g$ and in particular ${\cal D/N}=1$ above and below
$T_g$. Thus, it seems that the theory of \cite{BG} does not apply to our model
below $T_g$.
Note however that this theory is developed under the assumption of
a strong chaoticity of the system, while below $T_g$ and for
$\gamma \sim 0$ the dynamics of our system becomes slower and slower.
Thus, our results should not be regarded as invalidating the conjecture of
\cite{BG}, but as indicating that the hypothesis of \cite{BG} (essentially,
the requirement of strong chaoticity) are not satisfied by our model below $T_g$.

To resume, we studied the fluctuations of entropy production in a numerical
simulation of a Lennard-Jones like fluid above and below the glass transition
temperature $T_g$. We showed that below $T_g$ the Fluctuation Relation does not
hold; in particular, our data are consistent with a modified form of the FR
expressed by Eq.~\ref{FTX}.
We also showed that the behavior of the temperature derived from
Eq.~\ref{FTX}, $T_{FR} = T/X$, is qualitatively similar to that of
the effective temperature $T_{eff}$ that is usually defined from the 
generalized Fluctuation Dissipation Relation.
A relation between $T_{FR}$ and $T_{eff}$ has been proposed in 
\cite{FRgen-vetri,semerjian}
and our result are consistent with a recent quantitative derivation of
this relation in a simplified model \cite{CKZ04}.
Finally, we tested a conjecture that relates the factor $X$ in Eq.~\ref{FTX}
to properties of the phase space of the considered system; unfortunately, our
data are not consistent with this conjecture; thus, we believe that the
violation of the FR is, in our case, of different origin than that proposed
in \cite{BG}. We hope that future work will clarify this important issue.

We are indebted to L.~Cugliandolo, J.~Kurchan, 
G.~Gallavotti, A.~Giuliani and G.~Parisi for stimulating discussions.
We also thank C.~De~Michele, G.~Foffi and S.~Mariossi for their invaluable help 
in operating the FDT cluster of the INFM-CRS Soft 
on which the numerical computations have been performed.

\begin{thebibliography}{99}

\bibitem{EGD} D.~J.~Evans, E.~G.~D.~Cohen, and G.~P.~Morriss,
Phys.~Rev.~Lett. {\bf 71}, 2401 (1993).

\bibitem{GC} G.~Gallavotti and E.G.D.~Cohen,
Phys.~Rev.~Lett. {\bf 74}, 2694 (1995);
G.~Gallavotti, Mathematical Physics Electronic Journal {\bf 1}, 1 (1995).

\bibitem{GAL} G.~Gallavotti,
Phys.~Rev.~Lett. {\bf 77}, 4334 (1996).

\bibitem{BG}  F.~Bonetto, G.~Gallavotti, and P.~L.~Garrido,
Physica D {\bf 105}, 226 (1997);
G.~Gallavotti, Open Systems and Information Dynamics {\bf 6}, 101 (1999).

\bibitem{FTgen} 
F.~Bonetto, N.~I.~Chernov, J.~L.~Lebowitz,
Chaos {\bf 8}, 823 (1998);
 L.~Biferale, D.~Pierotti, and A.~Vulpiani,
J.~Phys.~A:~Math.~Gen. {\bf 31}, 21 (1998);
J.~L.~Lebowitz and H.~Spohn,
J.~Stat.~Phys. {\bf 95}, 333 (1999);
L.~Rondoni and E.~Segre, Nonlinearity {\bf 12}, 1471 (1999).

\bibitem{Kurchan} J.~Kurchan,
J.~Phys.~A:~Math.~Gen. {\bf 31}, 3719 (1998).

\bibitem{noiJSP} F.~Zamponi, G.~Ruocco, and L.~Angelani,
J.~Stat.~Phys {\bf 115}, 1655 (2004).

\bibitem{FTexp} S.~Ciliberto and C.~Laroche,
J.~Phys.~IV {\bf 8}, 215 (1998);
 W.~I.~Goldburg, Y.~Y.~Goldschmidt, and H.~Kellay,
Phys.~Rev.~Lett. {\bf 87}, 245502 (2001);
 K.~Feitosa and N.~Menon, 
Phys.~Rev.~Lett. {\bf 92}, 164301 (2004);
S.~Ciliberto, N.~Garnier, S.~Hernandez, C.~Lacpatia, J.-F.~Pinton, 
and G.~Ruiz~Chavarria, Physica A {\bf 340}, 240 (2004).

\bibitem{drivenglassy}
J.L.~Barrat and L.~Berthier,
Phys.~Rev.~E {\bf 63}, 012503 (2000);
L.~Berthier and J.L.~Barrat,
J.~Chem.~Phys. {\bf 116}, 6228 (2002).

\bibitem{nota1} Note that in numerical simulations the glass transition
essentially coincides with the {\it mode-coupling temperature} $T_{MCT}$ while
in experiments $T_g < T_{MCT}$. 

\bibitem{FDTshear}
L.~F.~Cugliandolo and J.~Kurchan,
Phys.~Rev.~Lett. {\bf 71}, 173 (1993);
L.~F.~Cugliandolo, J.~Kurchan, and L.~Peliti,
Phys.~Rev.~E {\bf 55}, 3898 (1997);
 L.~Berthier, J.L.~Barrat, and J.~Kurchan,
Phys.~Rev.~E {\bf 61}, 5464 (2000).

\bibitem{gal1} G.~Gallavotti, Chaos, {\bf 14}, 680 (2004);
G.~Gallavotti and E.~G.~D.~Cohen, Phys.~Rev.~E {\bf 69}, 035104 (2004).

\bibitem{FRgen-vetri}
M.~Sellitto, cond-mat/9809186;
A.~Crisanti and F.~Ritort, Europhys.~Lett. {\bf 66}, 253 (2004).

\bibitem{semerjian}
G.~Semerjian, L.~F.~Cugliandolo, and A.~Montanari,
J.~Stat.~Phys {\bf 115}, 493 (2004).

\bibitem{CKZ04}
L.~F.~Cugliandolo, J.~Kurchan and F.~Zamponi, in preparation.

\bibitem{PhDcristiano}
C.~De~Michele, F.~Sciortino, and A.~Coniglio,
J.~Phys.:~Condens.~Matter {\bf 16}, L489 (2004);
L.~Angelani, C.~De~Michele, G.~Ruocco and F.~Sciortino,
J.~Chem.~Phys. {\bf 121}, 7533 (2004);
the constants $\epsilon_{\alpha\beta}$ and $\sigma_{\alpha\beta}$ are the
same of the LJ Kob-Andersen mixture, see Ref.~\cite{drivenglassy} and 
references therein.

\bibitem{evans} D.J.~Evans and G.P.~Morris, {\it Statistical Mechanics of
Nonequilibrium Liquids} (Academic, London, 1990).

\bibitem{nota2} In this paper we will not address the problem of the
identification of the entropy production rate with the phase space contraction
rate. This point is discussed in detail in \cite{noiJSP} for our model.

\bibitem{FDT-MSD} R.~Di~Leonardo, L.~Angelani, G.~Parisi, and G.~Ruocco,
Phys.~Rev.~Lett. {\bf 84}, 6054 (2000).

\bibitem{pairing} D.~J.~Searles, D.~J.~Evans, D.~J.~Isbister,
Chaos {\bf 8}, 337 (1998);
C.~P.~Dettmann and G.~P.~Morriss,
Phys.~Rev.~E {\bf 53}, R5545 (1996).

\bibitem{BGS76} G.~Benettin, L.~Galgani, J.-M.~Strelcyn, 
Phys.~Rev.~A {\bf 14}, 2338 (1976).

\end{thebibliography}
\end{document}